\PassOptionsToPackage{usenames,dvipsnames}{xcolor}
\documentclass[pdflatex,sn-nature,iicol]{sn-jnl}
\usepackage{graphicx}%
\usepackage{multirow}%
\usepackage{amsmath,amssymb,amsfonts}%
\usepackage{amsthm}%
\usepackage{mathrsfs}%
\usepackage[title]{appendix}%
\usepackage{xcolor}%
\usepackage{textcomp}%
\usepackage{manyfoot}%
\usepackage{algorithm}%
\usepackage{algorithmicx}%
\usepackage{algpseudocode}%
\usepackage{listings}%
\usepackage{booktabs}
\usepackage{svg}
\usepackage{caption}
\usepackage[utf8]{inputenc}
\usepackage{tabularx}
\usepackage{array}
\usepackage{booktabs}
\usepackage{url}
\usepackage[version=4]{mhchem}
\usepackage{xcolor}
\usepackage{xspace}
\usepackage{soul}
\usepackage{titlesec}
\usepackage[normalem]{ulem}

\titleformat{\subsection}
  {\large\bfseries}
  {}
  {0pt}
  {}

\titleformat{\subsubsection}
  {\normalsize\bfseries}
  {}
  {0pt}
  {}

\titlespacing*{\subsection}
  {0pt}{2.0ex plus 0.5ex minus 0.2ex}{1.0ex}

\titlespacing*{\subsubsection}
  {0pt}{1.3ex plus 0.4ex minus 0.2ex}{0.5ex}

\sethlcolor{yellow} 

\theoremstyle{thmstyleone}%
\theoremstyle{thmstyletwo}%

\theoremstyle{thmstylethree}%
\DeclareRobustCommand{\xrdreader}{ERAF4{\mdseries\textsc{xrd}}\xspace}
\graphicspath{ {./figures/} }

\definecolor{dgreen}{HTML}{008000}

\definecolor{codegreen}{rgb}{0,0.6,0}
\definecolor{codegray}{rgb}{0.5,0.5,0.5}
\definecolor{codepurple}{rgb}{0.58,0,0.82}
\definecolor{backcolour}{rgb}{0.95,0.95,0.92}
\lstdefinestyle{mystyle}{
    backgroundcolor=\color{backcolour},
    commentstyle=\color{codegreen},
    keywordstyle=\color{magenta},
    numberstyle=\tiny\color{codegray},
    stringstyle=\color{codepurple},
    basicstyle=\ttfamily\footnotesize,
    breakatwhitespace=false,
    breaklines=true,
    captionpos=b,
    keepspaces=true,
    numbers=left,
    numbersep=5pt,
    showspaces=false,
    showstringspaces=false,
    showtabs=false,
    tabsize=2
}
% ...
\newcounter{saveenumi}

\newcommand{\be}{\begin{enumerate}}
\newcommand{\ee}{\end{enumerate}}
\newcommand{\bes}{\begin{enumerate}[wide, labelwidth=!, labelindent=0pt, label=\textbf{{\color{blue}\arabic*}.}]}
\newcommand{\ees}{\end{enumerate}}

\title{\xrdreader: A multimodal agentic framework for constructing validated experimental X-ray diffraction databases from scientific literature}

\newcommand{\authorfontsize}{\fontsize{9}{11}\selectfont}
\newcommand{\affilfontsize}{\fontsize{8}{10}\selectfont}

\author[1]{{\authorfontsize \fnm{Afnan} \sur{Mostafa}}}

\author[2,3]{{\authorfontsize \fnm{William} \sur{Ratcliff}}}

\author[4]{{\authorfontsize \fnm{Simon~J.~L.} \sur{Billinge}}}

\author*[1,5]{{\authorfontsize \fnm{Niaz} \sur{Abdolrahim}}}\email{niaz@rochester.edu}

\affil[1]{{\affilfontsize \orgdiv{Department of Mechanical Engineering}, \orgname{University of Rochester}, \orgaddress{\postcode{NY 14627}, \country{USA}}}}

\affil[2]{{\affilfontsize \orgdiv{NIST Center for Neutron Research}, \orgname{National Institute of Standards and Technology, Gaithersburg}, \orgaddress{\postcode{MD 20899}, \country{USA}}}}

\affil[3]{{\affilfontsize \orgdiv{Department of Physics, Department of Materials Science and Engineering}, \orgname{University of Maryland, College Park}, \orgaddress{\postcode{MD 20742}, \country{USA}}}}

\affil[4]{{\affilfontsize \orgdiv{Department of Materials}, \orgname{University of California, Santa Barbara}, \orgaddress{\postcode{CA 93106}, \country{USA}}}}

\affil[5]{{\affilfontsize \orgdiv{Laboratory for Laser Energetics}, \orgname{University of Rochester}, \orgaddress{\postcode{NY 14623}, \country{USA}}}}

\date{\today}

\begin{document}

\abstract{
The scientific literature contains decades of experimental measurements that remain difficult to access as structured data for modern AI and data-driven research. Much of this information is distributed across figures, captions, text, and tables, requiring experimental data and their context to be identified, connected, and verified before they can be reused. Here we introduce \xrdreader (Experiment Reader Agentic Framework for X-Ray Diffraction), a fully automated multimodal (\textit{i.e.}, image and text), multi-agent framework that reconstructs validated X-ray diffraction (XRD) records from scientific publications. \xrdreader downloads and screens documents, identifies XRD figures, extracts and links metadata to the corresponding experimental data, and validates outputs against source evidence using an independent validation agent. On a manually curated benchmark of 273 scientific publications containing 3,150 candidate figures, \xrdreader achieved up to 98.7\% accuracy for XRD figure identification and generated 1,400 metadata values across 22 fields. Independent manual assessment of the final validated records yielded 98.5\% precision and 90.7\% recall, with no unsupported metadata observed among 443 evaluated fields. By moving beyond information extraction to the reconstruction and validation of linked experimental records, \xrdreader establishes an automated approach for transforming published scientific information into machine-readable experimental datasets for AI and data-driven science.}

\keywords{literature mining, powder X-ray diffraction, large language models, agentic workflows, machine-readable databases}

\maketitle


Modern artificial intelligence (AI) and machine-learning (ML) workflows rely on large quantities of structured, annotated, and machine-readable data~\cite{himanen2019data,choudhary2022recent}. 
In materials science, such datasets underpin applications ranging from automated characterization and property prediction to data-driven materials discovery~\cite{goodall2020predicting,salgado2023automated,liu2019using,gomez2023convolutional,guo2024towards,hetti2026machine,szymanski2023adaptively,guo2025ab,shahnazari2026machine,kjaer2023deepstruc,kusne2020fly}.
A substantial fraction of scientific data resides in publications designed for human reading rather than computational reuse~\cite{olivetti2020data,stocker2025rethinking}. 
Scientific results and their context are distributed across figures, captions, text, tables, and supplementary information~\cite{himanen2019data,hollarek2026opxrd}, leaving much of the published scientific record inaccessible to modern AI and ML workflows.
Converting even a fraction of this record into structured datasets could substantially expand the experimental data available for AI-driven science.

Experimental data are frequently reported graphically while the underlying numerical data are unavailable~\cite{kroon2017raw,kroon2024raw}. 
Unlocking this information, therefore, requires approaches that recover not only the experimental data, but also the contextual metadata necessary to interpret them, and organize both into structured, machine-readable records~\cite{wilkinson2016fair}.
This task is challenging because the contextual metadata associated with an experimental dataset is rarely colocated with the data in the manuscript~\cite{himanen2019data,hollarek2026opxrd,khalighinejad2025matvix}. 
Figures that are relevant must first be identified, then interpreted, including the quantities and units represented, while complementary information elsewhere in the publication—such as sample composition, measurement conditions, or experimental configuration—must be extracted and correctly linked with the data in that figure. 
Establishing these data--metadata links is essential. For example, a metadata value may be correct in isolation yet incorrect for a particular experimental record if linked to the wrong figure. 

Recent advances in large language models (LLMs), vision-language models (VLMs), and agentic AI provide new capabilities for extracting and interpreting information distributed across scientific text and figures and transforming it into structured records~\cite{polak2024extracting,rameshbabu2026papers,li2024multimodal,odobesku2025agent,zhang2026towards,kang2024chatmof}.
Agentic workflows are particularly promising because models can retrieve evidence, use external tools, and iteratively resolve information rather than relying on a single model response.

For scientific database construction, however, extraction alone is insufficient. Errors in the extracted data, metadata, or data--metadata links can rapidly weaken the reliability of the resulting database and the downstream AI and ML workflows~\cite{wei2025ai,huang2025survey}.
Moreover, validation itself must be automated if literature-to-database workflows are to operate at scale without human intervention. 
A fully automated framework must therefore not only extract scientific information but also validate the reliability of its own outputs against the original source evidence, enabling the construction of machine-readable records at scale without manual curation.

Here, we demonstrate \xrdreader (\textbf{E}xperiment \textbf{R}eader \textbf{A}gentic \textbf{F}ramework for \textbf{X}-\textbf{R}ay \textbf{D}iffraction), an automated framework for the construction of experimental records from the scientific literature using experimental XRD, one of the most widely used techniques for structural characterization~\cite{callister2020materials,newnham2004properties,kalidindi2015materials}.
XRD provides a demanding test case: increasing experimental automation is generating diffraction measurements at scales that motivate automated interpretation~\cite{lee2020deep,shahzad2024accelerating,cawse2001experimental}, while decades of published diffraction measurements constitute a large resource of expert-interpreted experimental information. 
Yet XRD patterns are commonly embedded as figures, with the metadata required for their interpretation distributed throughout the publication~\cite{himanen2019data,hollarek2026opxrd}.

Scientific literature mining has progressed from rule- and natural language processing-based tools such as OSCAR~\cite{jessop2011oscar4} and ChemDataExtractor~\cite{swain2016chemdataextractor} to domain-specific language models, such as MatScholar~\cite{weston2019named} and MaterialsBERT~\cite{shetty2023general}, as well as other models~\cite{mikolov2013distributed,devlin2019bert,gupta2022matscibert}, and more recently, LLM-based scientific information extraction~\cite{dagdelen2024structured,polak2024extracting,rameshbabu2026papers}.
Related developments include crystallography- and diffraction-focused models~\cite{antunes2024crystal,vosoughi2026openxrd,johansen2025decifer,choudhary2025diffractgpt} and agentic materials workflows that combine language models with iterative tool use, such as nanoMINER~\cite{odobesku2025agent}, DIVE~\cite{zhang2026dive}, and other approaches~\cite{chen2025chemminer,li2025slm}.
In parallel, opXRD has demonstrated the potential of large, open-access experimental XRD databases by aggregating diffraction data from existing data repositories and contributions shared by researchers~\cite{hollarek2026opxrd}.
These efforts provide valuable experimental data resources; however, \xrdreader addresses a much larger challenge of recovering experimental XRD data and associated metadata embedded within the scientific literature, which is not readily accessible in machine-readable form.
Although demonstrated here for XRD, the framework can be adapted to other scientific domains where multimodal extraction and data--metadata linking are critical, supporting a broader literature-to-database concept for downstream automated AI- and ML-driven scientific workflows.

\begin{figure*}
\begin{center}
\includegraphics[scale=0.385]{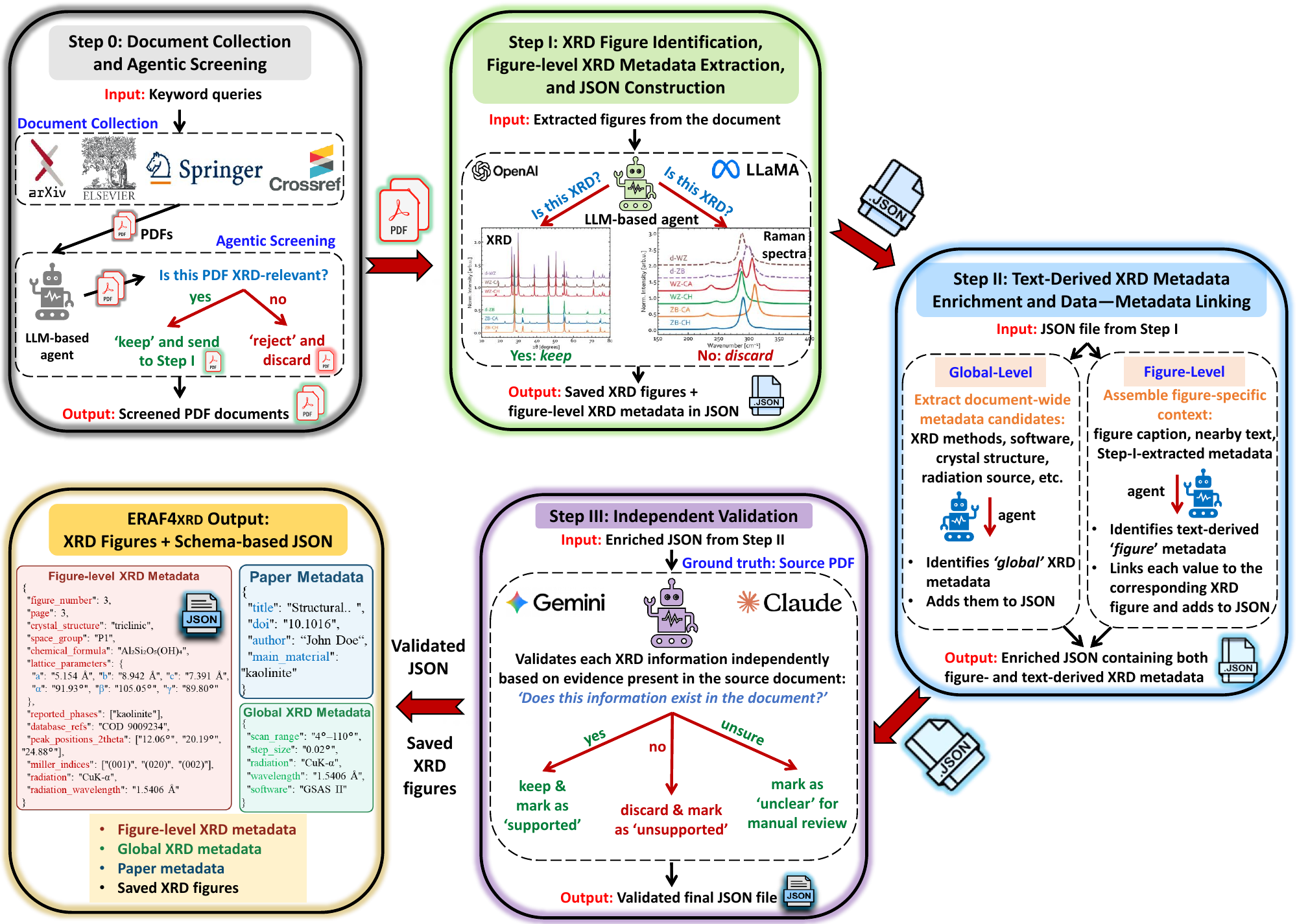}
\caption{\label{fig:framework}
\xrdreader framework for automated extraction, validation, and structuring of XRD figures and metadata from scientific literature. The framework consists of four steps. Step~0 collects open-access PDF documents and screens them for XRD content. For retained documents, Step~I extracts figure candidates, identifies and saves confirmed XRD figures, extracts figure-level XRD metadata, and constructs the initial JSON file; the example XRD and Raman figures are adapted from Ref.~\cite{larsen2022experimental}. Step~II enriches the JSON with text-derived figure-level metadata and global XRD methodological metadata. Step~III independently validates XRD information against the source document and produces the final validated JSON output. The final output contains XRD figures, paper metadata, XRD figure-level metadata, and global XRD methodological metadata.
}
\end{center}
\end{figure*}

Generally, a complete literature-to-database workflow requires several tasks: document collection and screening, figure identification, extraction of figure- and text-derived metadata, data--metadata linking, validation of the resulting records, and ultimately recovery of the numerical data contained within figures.
The current implementation of \xrdreader performs all these tasks without human intervention, except for recovering numerical data from figures.
For literature-scale deployment, these tasks must be performed automatically and consistently, without requiring human validation for individual records. 
We therefore treat validation not as a manual post-processing step but as an automated integral component of the extraction architecture; we refer to this as automated validation.

We rigorously assess the performance of \xrdreader against extensive human-labeled benchmark datasets spanning document screening, XRD figure identification, metadata extraction, and data--metadata linking; we refer to this as manual assessment. 
For this purpose, we created a benchmark dataset of 273 publications containing 282 manually verified XRD figures.
The publications were inspected by two reviewers to establish ground truth for XRD content, total figure count, and XRD figure count.
Note that this manual assessment is not part of the framework and is used only to evaluate the automated validation. We make these ground-truth datasets available publicly so they may also provide benchmarks for future scientific information-extraction workflows.

\begin{figure*}
\begin{center}
\includegraphics[scale=0.25]{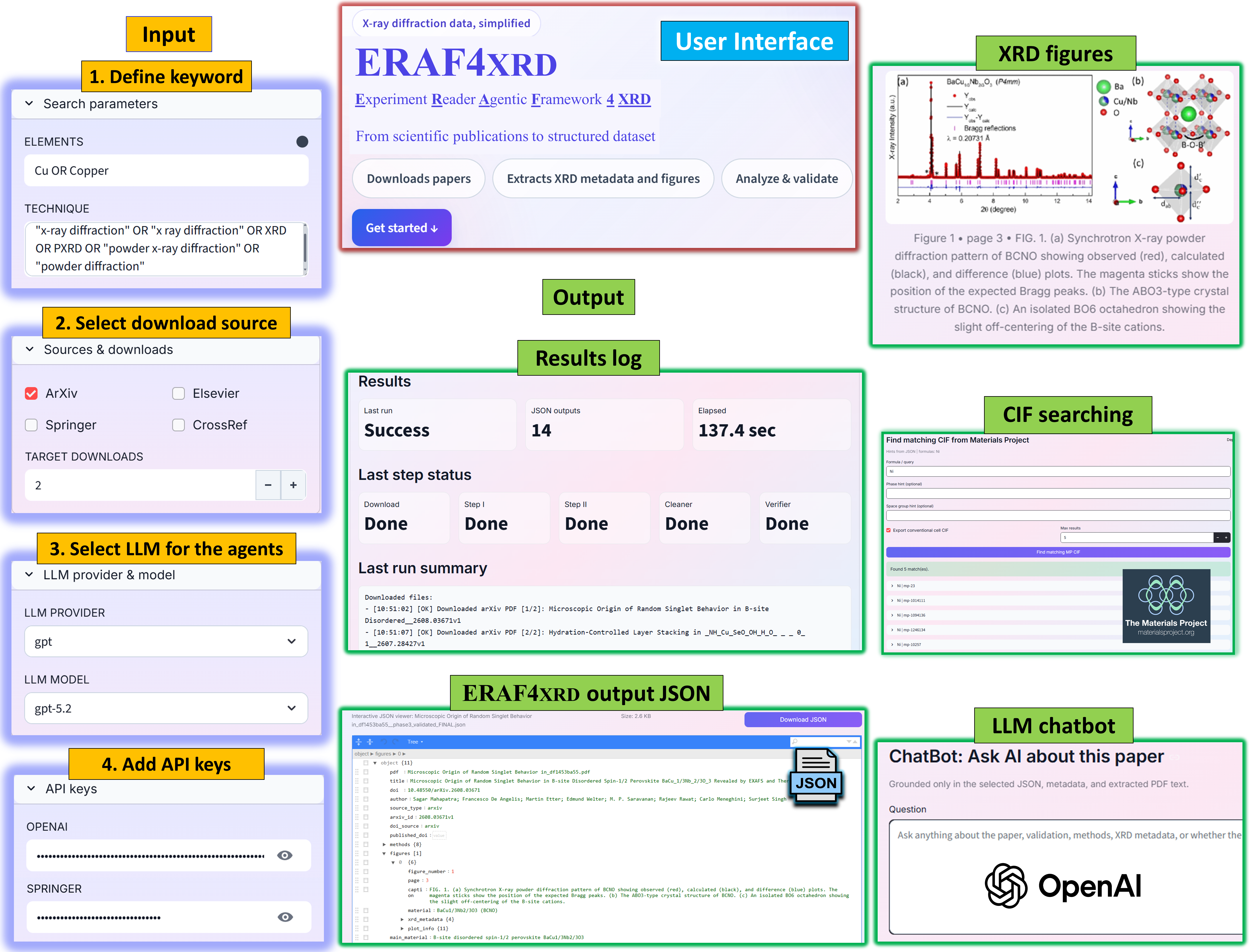}
\caption{\label{fig:ui} \xrdreader user interface. Blue-outlined panels show user inputs and configuration options, including keyword search parameters, download sources, LLM settings, and API keys. Green-outlined panels correspond to outputs, including logs, JSON results, \xrdreader-identified XRD plots, Materials Project CIF matches \cite{jain2013commentary}, and the integrated LLM chatbot. The XRD figure is taken from~\cite{mahapatra2026microscopic}.}
\end{center}
\end{figure*}

We further compare multiple LLMs and strategies, including agentic versus single-pass modes, targeted preprocessing versus full-document input, and different LLM visual-input configurations for figure identification (\textit{i.e.}, figure plus full page, figure only, or full page only) and introduce the best approaches for recovery, correct linking, and automated validation.
These experiments quantify trade-offs among accuracy, runtime, token consumption, and API cost and inform the current design of \xrdreader.
For example, we identify when single-pass mode can match the performance of agentic mode and when agentic reasoning provides a measurable advantage.

\begin{table*}
\centering
\caption{Benchmark dataset used for \xrdreader performance evaluation.}
\label{tab:benchmark_datasets}
\renewcommand{\arraystretch}{1.1}
\footnotesize
\begin{tabularx}{\textwidth}{c >{\raggedright\arraybackslash}X >{\raggedright\arraybackslash}X c c c c c c}
\toprule
\# & Material & Source & PDFs & Figures & Candidates & XRD & XRD (\%) & Candidate/XRD \\
\midrule
1 & Generic   & ArXiv    & 50 & 482 & 643 & 75 & 11.66 & 8.57 \\
2 & HEA/MEA   & ArXiv    & 50 & 540 & 667 & 15 & 2.25  & 44.47 \\
3 & Fe or Cu  & ArXiv    & 51 & 417 & 522 & 10 & 1.92  & 52.20 \\
4 & Al alloys & ArXiv    & 50 & 492 & 572 & 22 & 3.85  & 26.00 \\
5 & Mo        & Elsevier & 16 & 157 & 174 & 34 & 19.54 & 5.12 \\
6 & Fe or Cu  & Elsevier & 15 & 142 & 206 & 34 & 16.50 & 6.06 \\
7 & Oxides    & CrossRef & 22 & 144 & 219 & 64 & 29.22 & 3.42 \\
8 & Fe or Cu  & Springer & 19 & 151 & 147 & 28 & 19.05 & 5.25 \\
\midrule
& \multicolumn{2}{c}{} & \textbf{273} & \textbf{2525} & \textbf{3150} & \textbf{282} & \textbf{8.95} & \textbf{11.17} \\
\bottomrule
\end{tabularx}
\end{table*}

\xrdreader operates in four steps to automatically collect, extract, link, and validate experimental XRD information from publications (Fig.~\ref{fig:framework}).
Step~0 collects open-access publications in portable document format (PDF) from ArXiv, Elsevier, Springer, and CrossRef using user-defined keyword queries and screens them for XRD content. 
Step~I then extracts candidate figures, identifies XRD figures using a vision-capable LLM, and constructs per-publication JSON records containing figure-derived metadata. 
Step~II enriches these records with text-derived metadata and links them to the corresponding XRD figures. 
Finally, Step~III validates the XRD figures, metadata, and data--metadata links against source evidence before producing the final structured outputs.
Details of each step are provided in the Methods section and in Supplementary Notes 1-4.
\xrdreader is also accessible through a local graphical user interface (Fig.~\ref{fig:ui}) that enables researchers to process their own document collections and search for literature-derived XRD information.

The results are organized into six parts: benchmark dataset, reconstruction and automated validation of experimental records, manual assessment of the reliability of \xrdreader outputs, strategies for successful experimental record reconstruction, analysis of failed cases in XRD figure identification and metadata extraction, and scalability in terms of runtime, cost, resource--accuracy trade-offs, and LLM selection.
\xrdreader supports multiple LLM providers, including GPT, Gemini, Claude, and Grok; unless otherwise stated, the results presented were obtained using \texttt{GPT-5.2}.

\section*{Results}


\phantomsection
\subsection*{Benchmark dataset}\label{subsec:benchmark}

We evaluated \xrdreader using a benchmark dataset that comprises 273 open-access and Creative Commons-licensed publications spanning pure metals, oxides, alloys, and one material-independent set (Table~\ref{tab:benchmark_datasets}).
These publications contain 2,525 figures, while \xrdreader's Step~I extracted 3,150 candidate figures because non-figure entities (\textit{e.g.}, journal logos) were also captured during candidate figure extraction (see Methods and Supplementary Discussion 1).
Of these candidates, 282 were manually verified as true XRD figures, corresponding to a candidate-to-XRD ratio of approximately 11:1 (\textit{i.e.}, one true XRD figure for every 11 candidates) and creating an imbalanced classification task.
A similar imbalance (100:1) has been reported for textual materials-data extraction~\cite{polak2024extracting}.
Supplementary Discussion 1 analyzes candidate-to-XRD ratios across individual datasets, identifying those with relatively low ratios. 
It also shows the publication-size distribution, demonstrating that \xrdreader was evaluated on both short and long documents (Supplementary Fig.~3), with longer documents posing greater challenges for contextual understanding~\cite{liu2024lost}. Ground truth was established through manual review, with additional benchmark subsets used to evaluate metadata extraction, data--metadata linking, and alternative extraction strategies.



\begin{figure}
  \centering
  \includegraphics[scale=0.35]{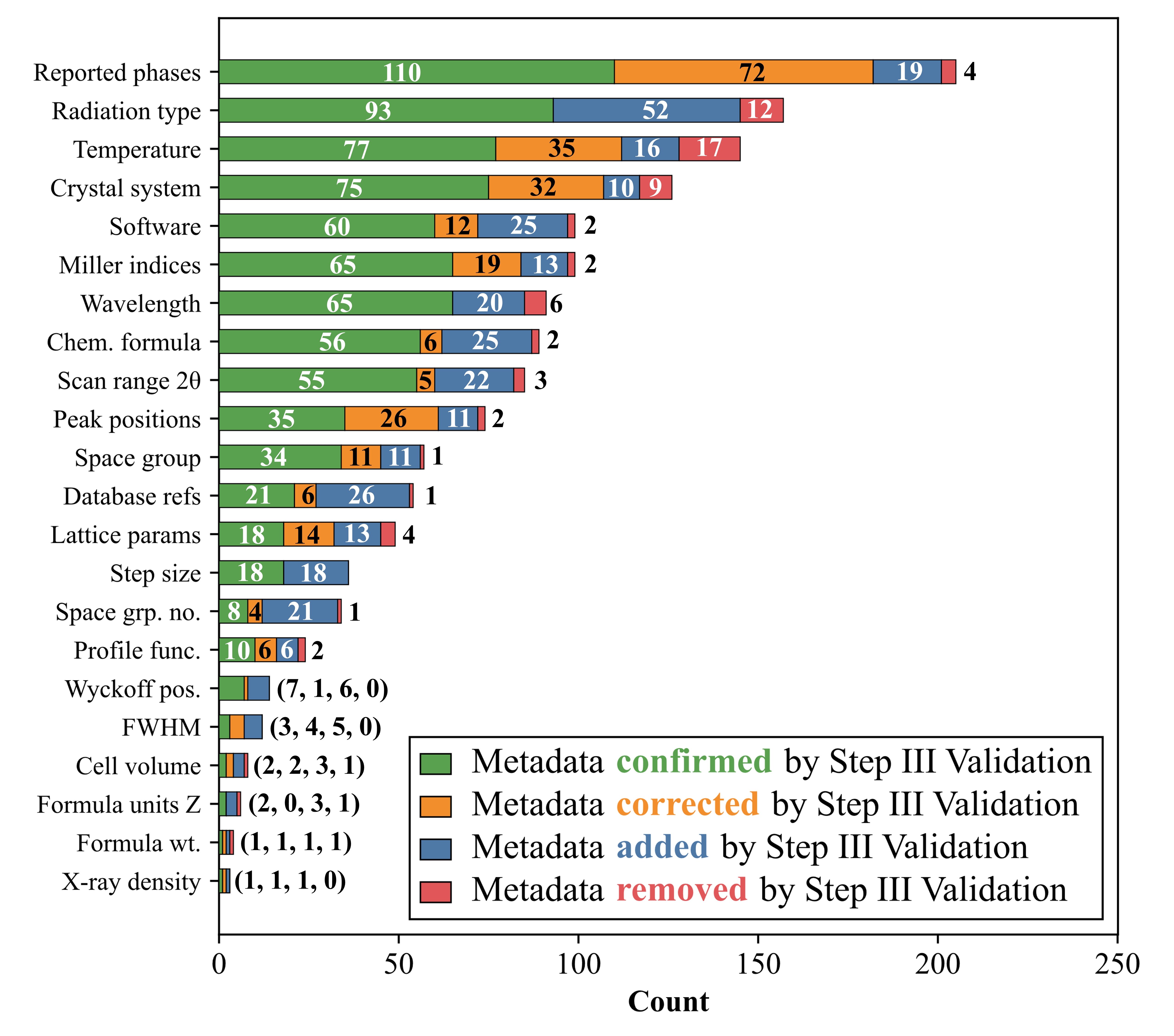}
  \caption{Step~III automated validation outcomes for XRD metadata across the full benchmark dataset using \texttt{GPT-5.2}. Each horizontal stacked bar shows the number of XRD metadata values confirmed (green), corrected (orange), newly added (blue), or removed (red) by Step~III validation. Values within or near the bars indicate the corresponding counts; for low-count fields, the values are listed to the right in the order confirmed, corrected, added, and removed.}
  \label{fig:stepIII_metadata_validation}
\end{figure}

\subsection*{\xrdreader reconstructs and automatically validates experimental records}
The central challenge is constructing reliable experimental records by linking fragmented metadata to the correct experimental data, particularly when publications contain multiple XRD figures.
\xrdreader addresses this challenge through an automated, source-grounded validation process that independently checks the extracted figures, metadata, and data--metadata links against evidence in the original publication, and corrects, adds, or removes information as needed.  
\xrdreader's Step~III independently validates the extracted metadata values and where they apply: figure-level metadata are retained only when the source document supports their link to the XRD figure, whereas global metadata apply across the document.
Validation proceeds through two sequential processes to identify from unsupported to inconsistent records: LLM-based source-evidence validation followed by non-LLM deterministic checks (see Methods).

Across 1,471 metadata values in the 273-publication benchmark, Step~III confirmed 816 values (55.5\%), and modified the remaining 655 (44.5\%) by correcting 257 (17.5\%), adding 327 (22.2\%), and removing 71 (4.8\%) (Fig.~\ref{fig:stepIII_metadata_validation}).
These results show that independent validation substantially changed the initial extracted results and demonstrate the importance of separated extraction and validation tasks. 

Of the 71 removed values, LLM-based validation eliminated 59. 44 values were not reported in the source document (\textit{i.e.}, hallucinations), 11 could not be linked to the correct XRD figure (\textit{i.e.}, incorrect data--metadata links), and 4 were contradicted by the source (Supplementary Fig.~4).
The remaining 12 were removed by deterministic checks, including 11 radiation--wavelength mismatches and one lattice-parameter--crystal-system mismatch.
Together, these complementary validation strategies prevent unsupported or hallucinated information from entering the final database, improving the reliability of the resulting records (Supplementary Discussion 2).

\begin{figure}
\begin{center}
\includegraphics[scale=0.22]{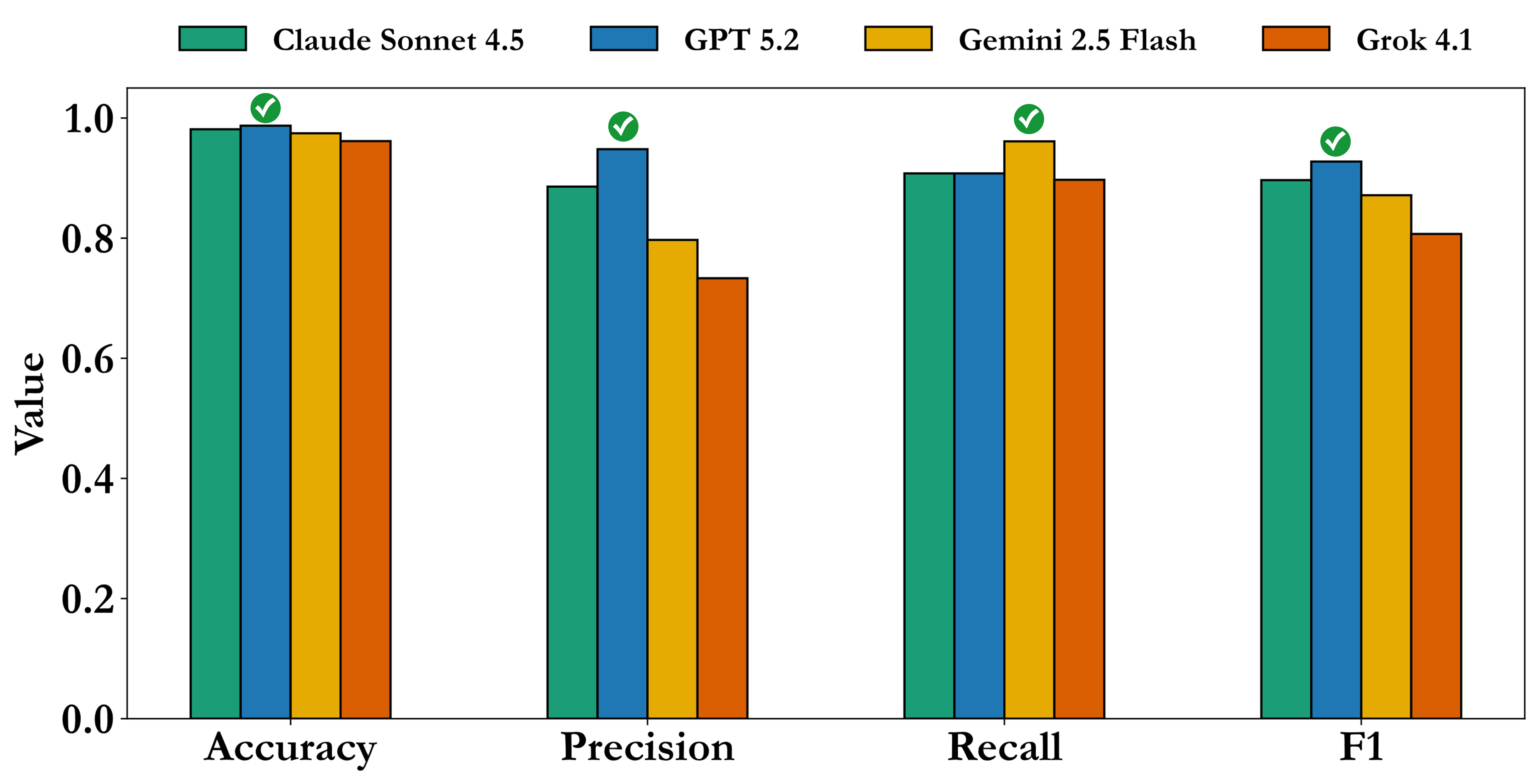}
\caption{\label{fig:metrics} XRD figure identification performance metrics across the full benchmark dataset. Accuracy, precision, recall, and F1-score are shown for \texttt{Claude Sonnet~4.5} (green), \texttt{GPT-5.2} (blue), \texttt{Gemini~2.5 Flash} (mustard yellow), and \texttt{Grok~4.1} (orange). Check marks indicate the best-performing model for each metric.
}
\end{center}
\end{figure}

Step~III also validates the XRD figures identified in Step~I through a low-cost, metadata-based consistency check without requiring LLMs for further vision-based identification process (see Methods). 
Each figure is evaluated for support from validated figure-specific or applicable global metadata and is flagged as a potential misidentification when neither is present.
Of the XRD figures identified using \texttt{GPT-5.2}, 81.8\% were supported by figure-level metadata and 17.9\% by applicable global metadata, and 0.4\% lacked any validated metadata and were flagged for review.

\subsection*{Manual assessment confirms the reliability of \xrdreader}

\textbf{XRD figure identification.} To independently assess the reliability of the automated validation, the final \xrdreader outputs are assessed against manually curated reference data. 
Independent manual assessment was used only to evaluate the final automated outputs and was not part of the \xrdreader framework. \xrdreader can reliably validate XRD figures, metadata, and data--metadata links without requiring any human validation.
For the XRD figure identification task, selected figures were compared with ground-truth XRD figures from the 273-publication benchmark dataset.
Four LLMs were tested in independent \xrdreader runs: \texttt{GPT-5.2}, \texttt{Gemini 2.5 Flash}, \texttt{Claude Sonnet 4.5}, and \texttt{Grok 4.1}.
The results are reported using accuracy, precision, recall, and F1-score, as shown in Fig.~\ref{fig:metrics}.
Accuracy ($\text{\textbf{A}}$) is defined as
\[
\text{\textbf{A}} = \frac{\text{TP} + \text{TN}}{\text{TP} + \text{TN} + \text{FP} + \text{FN}},
\]
where TP, TN, FP, and FN denote true positives, true negatives, false positives, and false negatives, respectively.
Here, TP and TN represent correctly identified XRD and non-XRD figures, respectively, while FP represents non-XRD figures incorrectly identified as XRD, and FN represents true XRD figures that were missed.

\begin{table*}
\centering
\caption{Illustrative examples of the metadata scoring outcomes.}
\label{tab:outcome_examples}

\begin{tabular*}{\textwidth}{@{\extracolsep{\fill}}lll@{}}
\toprule
Outcome & Source reported & \xrdreader{} output \\
\midrule
\emph{match}    & Cu K$\alpha$ & Cu K$\alpha$/Copper K$\alpha$ \\
\emph{incorrect} & Cu K$\alpha$ & Mo K$\alpha$ \\
\emph{miss}     & Cu K$\alpha$ & not extracted \\
\emph{unsupported} (LLM hallucination)    & not reported & Cu K$\alpha$ \\
\hline
true negative (unscored)   & not reported & not extracted \\
\bottomrule
\end{tabular*}
\end{table*}

\texttt{GPT-5.2} achieved the highest accuracy (98.7\%), followed by \texttt{Claude Sonnet 4.5} (98.1\%), \texttt{Gemini 2.5 Flash} (97.5\%), and \texttt{Grok 4.1} (96.2\%).
However, accuracy alone is insufficient to fully evaluate performance for two reasons.
First, the accuracy differences among the LLMs are small and may reflect benchmark-level statistical noise. Second, because the candidate pool is strongly imbalanced toward non-XRD figures, accuracy is dominated by correctly rejected negatives and does not adequately capture XRD identification performance (Supplementary Fig.~5). 
Since \xrdreader's goal is to contribute to building experimental XRD databases, reliably identifying true XRD figures is more important. We therefore focus on precision, recall, and F1-score, which are metrics that exclude TN.
Precision, defined as $\text{\textbf{P}}=\text{TP}/(\text{TP}+\text{FP})$, measures how many predicted XRD figures are truly XRD; therefore, it reflects FP control.
Recall, defined as $\text{\textbf{R}}=\text{TP}/(\text{TP}+\text{FN})$, measures how many true XRD figures are successfully found; therefore, it reflects FN behavior.
The F1-score, $\text{\textbf{F1}}=2(\text{\textbf{P}}\times\text{\textbf{R}})/(\text{\textbf{P}}+\text{\textbf{R}})$, balances $\text{\textbf{P}}$ and $\text{\textbf{R}}$.

\begin{figure}
\begin{center}
\includegraphics[scale=0.34]{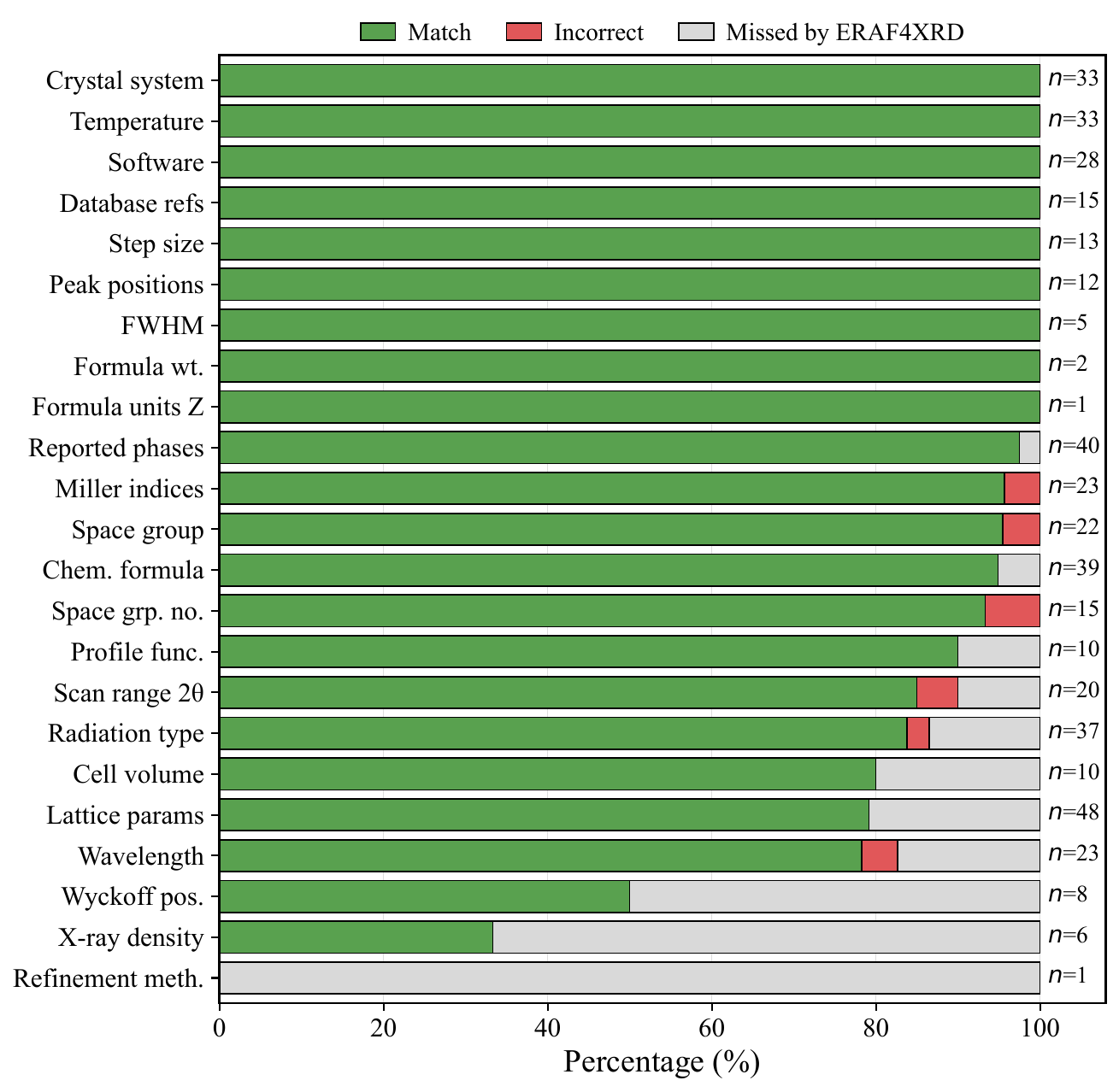}
\caption{\label{fig:metadata_perfield} Comparison of \xrdreader metadata outputs using \texttt{GPT-5.2} with manually curated reference metadata across the 47-publication subset. Each horizontal stacked bar shows the percentage of scored values that matched the reference metadata (green), disagreed with the reference metadata (red), or were missed by \xrdreader (gray). The total number of scored values ($n$) for each metadata field is shown on the right.}
\end{center}
\end{figure}

\texttt{GPT-5.2} presented the strongest XRD figure identification performance, with 94.8\% precision, 90.8\% recall, and a 92.8\% F1-score.
\texttt{Claude Sonnet 4.5} also showed similarly balanced performance (88.6\% precision, 90.8\% recall, and an 89.7\% F1-score), while
\texttt{Gemini 2.5 Flash} achieved the highest recall (96.1\%) but lower precision (79.7\%), indicating a more permissive identification behavior with more FP.
\texttt{Grok 4.1} maintained high recall (89.7\%) but had the lowest precision (73.3\%) and F1-score (80.7\%).
Overall, \texttt{GPT-5.2} showed the strongest performance, combining the highest F1-score with well-balanced precision and recall.
These metrics support the reliability of \xrdreader's automated XRD figure identification task.

Dataset-level analysis further showed that performance depends on both the LLM and publication corpus, with no model performing best across every dataset (Supplementary Discussion 3).
The results show that both dataset and LLM choices affect performance (Supplementary Figs.~6–9).
No single LLM performs best for every dataset, and some datasets, such as Fe/Cu--ArXiv, are challenging for all models.
The results also reveal distinct FP- and FN-dominated behavior (Supplementary Fig.~10); for example, with \texttt{GPT-5.2}, Fe/Cu--Elsevier was strongly FP-dominated, whereas Generic--ArXiv was FN-dominated.
This distinction is important because FP cases introduce incorrect records, whereas FNs leave true XRD figures absent, emphasizing reliability over completeness of \xrdreader.

\phantomsection\label{subsec:manual_metadata_assessment}
\textbf{Metadata extraction and data--metadata linking.} 
We next assessed the final validated metadata and data--metadata links against manually curated reference data from 47 publications. A metadata value was considered correct only when both its value and its association with the corresponding XRD figure agreed with the curated reference (Table~\ref{tab:outcome_examples}).

Across 443 evaluated metadata fields, 402 matched the curated reference, six were incorrect, 35 were missed, and no unsupported metadata were observed. This corresponds to 98.5\% precision and 90.7\% recall, showing that the final \xrdreader records are highly reliable when metadata are reported, with remaining errors dominated by missed rather than incorrect information.

Figure~\ref{fig:metadata_perfield} summarizes the assessment outcomes for \emph{match}es, \emph{incorrect} cases, and \emph{miss}es, revealing complete agreement for several metadata fields including crystal system, temperature, software, database references, step size, peak positions, FWHM, and formula weight.
Most misses occurred in lattice parameters, radiation type, wavelength, Wyckoff positions, and X-ray density.
Incorrect cases were less frequent and occurred for Miller indices, space group, space-group number, radiation type, wavelength, and scan range.
Notably, no unsupported metadata were observed, supporting the reliability of \xrdreader's validation in preventing hallucinated information from entering the final records.

\begin{figure*}
\begin{center}
\includegraphics[scale=0.38]{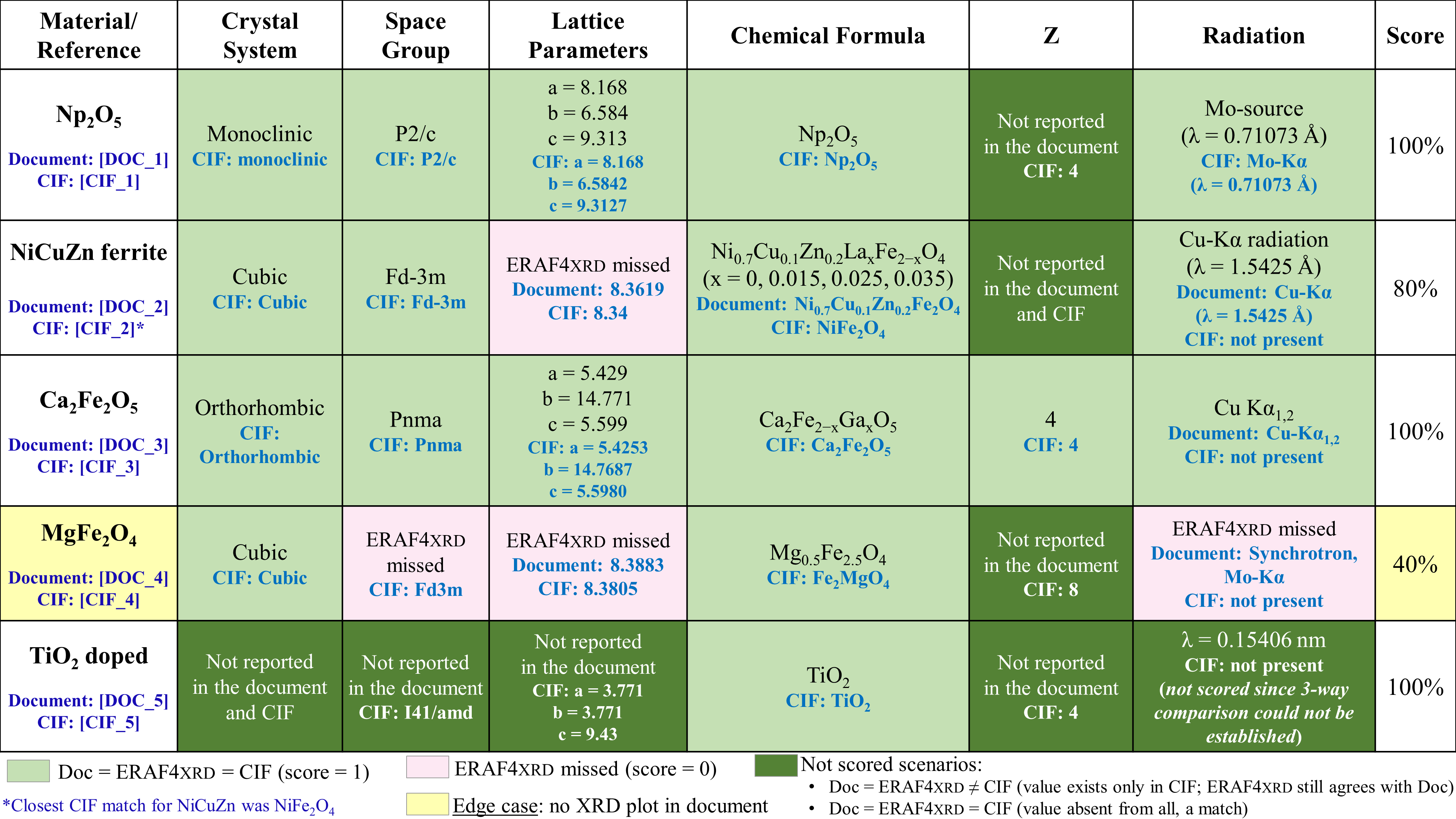}
\caption{\label{fig:metadata_3way} Three-way comparison of document-reported, \xrdreader-extracted (\texttt{GPT-5.2}), and CIF-derived XRD metadata. Light-green cells indicate agreement among all three sources when a value exists (score = 1), whereas pink cells indicate metadata present in both the document and CIF but missed by \xrdreader (score = 0). Dark-green cells indicate unscored cases in which the document and \xrdreader agree but either a value is present only in the CIF or no value is reported by any of the three sources. The yellow-highlighted `\ce{MgFe2O4}' row represents an edge case with no powder XRD diffractogram. References to the source documents and CIF files, labeled as \texttt{[DOC\_]} and \texttt{[CIF\_]}, respectively, are provided in Supplementary Table 3.}
\end{center}
\end{figure*}

\textbf{Comparing to CIF files.}
As a complementary assessment, we also compared \xrdreader-reported metadata with established crystallographic database CIF records from the Crystallography Open Database (COD)~\cite{gravzulis2009crystallography}.
For five randomly selected benchmark publications, metadata reported in the source publication and extracted by \xrdreader were compared with CIF-derived values for the corresponding material system. 
Across six crystallographic metadata fields—crystal system, space group, lattice parameters, chemical formula, formula units ($Z$), and radiation type (Fig.~\ref{fig:metadata_3way} and Supplementary Discussion 4), \xrdreader agreed with both the source publication and CIF reference for 18 of 22 scored fields (81.8\%).
After excluding one edge case without an extractable powder-XRD figure (yellow-highlighted cell), agreement increased to 16 of 17 fields (94.1\%).

Overall, the CIF-based assessment shows that \xrdreader not only recovers metadata reported in the source publications but also produces values that are highly consistent with established crystallographic records.

\phantomsection
\subsection*{Strategies for successful experimental record reconstruction}\label{subsec:design_choices}

We next benchmarked the strategies that determine \xrdreader performance and scalability. We evaluated alternative strategies for document screening, document-context selection provided to the LLM, visual inputs for XRD figure identification and data--metadata linking, and automated validation. Here, we summarize the key findings that guided the final \xrdreader design; detailed comparisons are provided in Supplementary Discussion 5.

\textbf{Document screening.}
Two screening strategies were identified: single-pass screening with a strong LLM for fast, low-cost high-throughput screening, and agentic screening with smaller models when additional evidence gathering was needed.
On a 35-publication subset, \texttt{GPT-5.2} achieved the same accuracy (97\%) in both modes, while single-pass screening used substantially fewer tokens and lower API costs (Supplementary Figs.~11–12).
In contrast, smaller models benefited from agentic reasoning, particularly for ambiguous cases where XRD was mentioned but extractable XRD information was absent.
For example, \texttt{GPT-4.1-nano} improved from 77\% to 89\% accuracy with an increased agentic budget.
Examples of agentic reasoning resolving ambiguous cases for smaller models are provided in Supplementary Tables~4-6.

\begin{figure*}
\begin{center}
\includegraphics[scale=0.45]{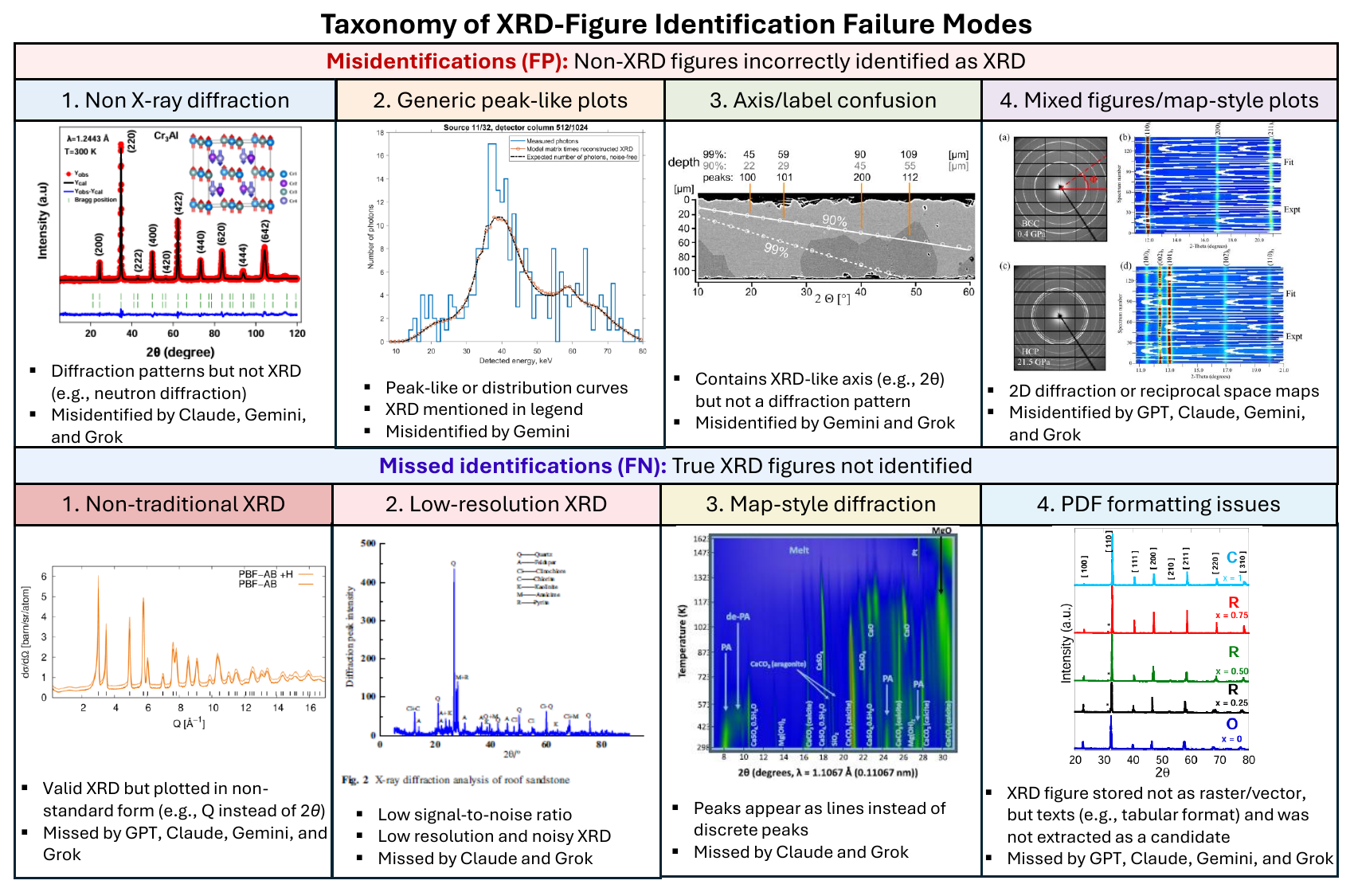}
\caption{\label{fig:taxonomy} Failure-mode taxonomy for \xrdreader Step~I XRD figure identification, with representative examples from each category. The top row shows misidentifications (FP): non X-ray diffraction, generic peak-like plots, axis/label confusion, and mixed figures/map-style plots (Refs.~\cite{philip2026disorder,korolkovas2022fast,wisniewski2023experimental,sahu2026pressure}, left to right). The bottom row shows missed cases (FN): non-traditional, low-resolution, map-style diffraction, and PDF formatting issues (Refs.~\cite{tappe2026impact,li2017mechanical,taira2024valorisation,erat2011electron}, left to right). The taxonomy was developed with the assistance of \texttt{GPT-5.4}, but all panels were manually curated from the published works.}
\end{center}
\end{figure*}

We next tested whether providing the complete publication improves screening relative to targeted preprocessing of XRD-relevant evidence. 
On the same 35-publication subset, \texttt{GPT-5.2} achieved higher accuracy with targeted pre-processed evidence than with full-document input (97\% versus 94\%), while requiring substantially fewer tokens and lower API cost (Supplementary Fig.~13). 
Increasing document context, \textit{i.e.}, uploading the full document, did not improve screening; instead, targeted preprocessing was both more accurate and more efficient.
Irrelevant text content, such as literature reviews, bibliographies, and acknowledgments, can dilute XRD-relevant evidence and consume limited context-window capacity, which is particularly important in agentic mode where sufficient space is needed for iterative evidence gathering.

\textbf{Figure identification and data--metadata linking.} For XRD figure identification and data--metadata linking, combining a high-resolution figure crop with the full-page view produced the strongest performance because the two inputs provide complementary plot-level and contextual information (Supplementary Figs.~14–15). This complementary visual context was particularly important for resolving data--metadata linking that could not be established reliably from either view alone, with the LLM's finite budgets~\cite{openai_vision_docs,anthropic_vision_docs,google_gemini_image_docs}, as illustrated in Supplementary Fig.~14.
Next, for metadata extraction and linking, regular-expression (regex)-based extraction followed by agentic LLM verification provided the strongest overall performance compared with direct LLM extraction and regex-based single-pass verification (Supplementary Table~7).

\textbf{Strategies for automated validation.}
To prevent misinformation from entering the resulting datasets, \xrdreader uses several validation strategies.
First, validation is performed after removing all prior agent traces accumulated during the preceding steps, allowing the validation agent to operate independently (Supplementary Fig.~1).
Second, \xrdreader supports the use of a different LLM to provide an additional layer of independent validation.
Third, the validation agent is the most resource-rich agent in the framework, with the largest decision budget and the broadest tool-use capabilities.
Fourth, the agent cannot retain a metadata value without providing supporting evidence from the source publication.
Finally, non-LLM deterministic checks evaluate the consistency of the retained metadata values. 
These checks are needed because \xrdreader treats the source publication as the ground truth for as-reported metadata. 
As a result, a value can exist in the source publication but still be inconsistent with basic crystallographic or radiation-source rules (see Methods).

\begin{figure}
\begin{center}
\includegraphics[scale=0.33]{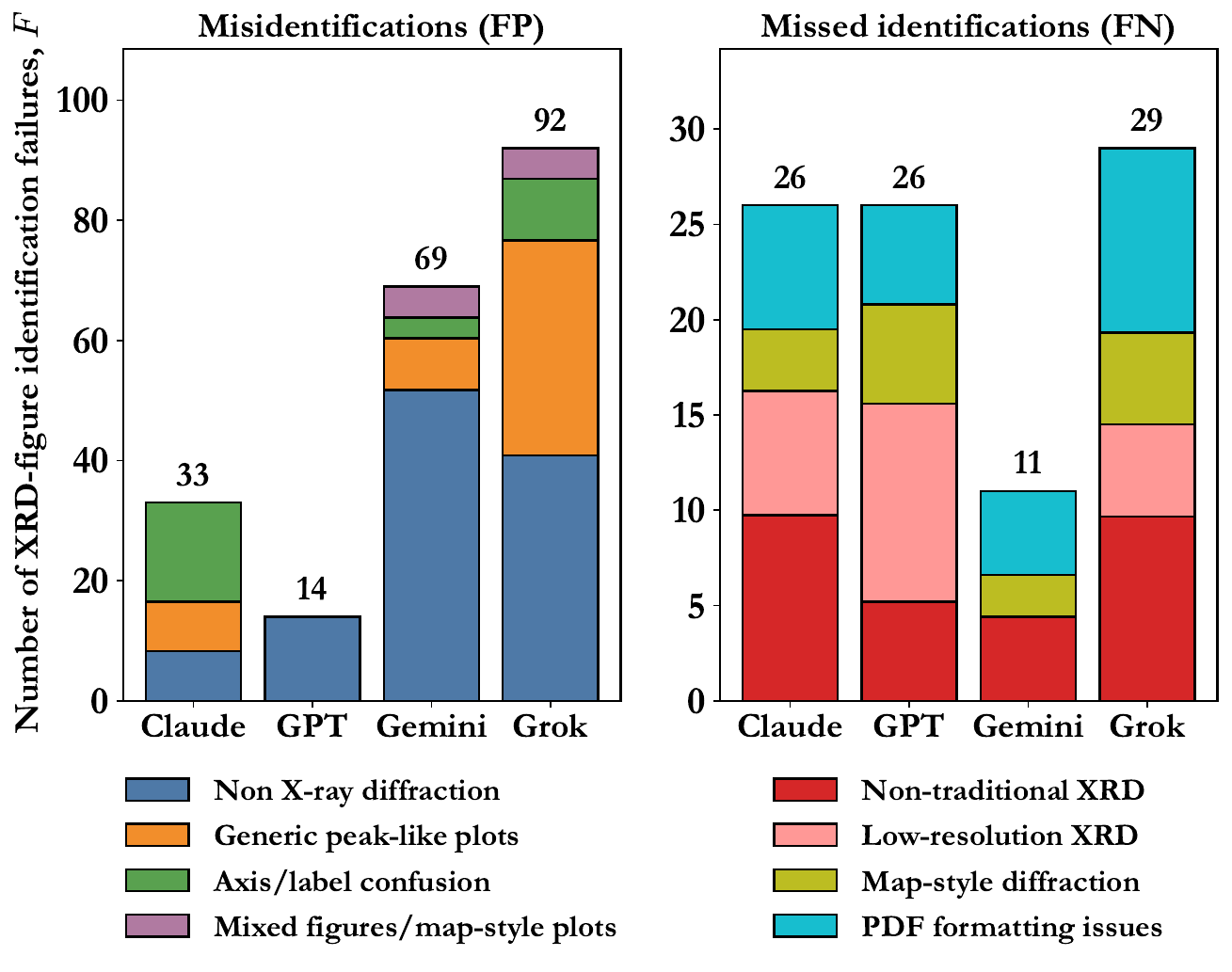}
\caption{\label{fig:failure_distribution}Distribution of XRD figure identification failures by taxonomy category. The left and right panels show misidentifications (FP) and missed identifications (FN), respectively. Numbers above the bars indicate the total FP or FN count for each model.}
\end{center}
\end{figure}

\subsection*{Analysis of failed cases}\label{subsec:failures}
The remaining failures revealed distinct patterns across tasks: metadata-extraction failures were dominated by missed information, whereas figure-identification failures more often involved misidentifications. 
Understanding these distinct failure modes is important because they affect database quality differently; incorrect values or links can introduce erroneous records, while omissions only reduce coverage.

\textbf{XRD figure identification failure modes.}
XRD figure identification failures were grouped into two types: misidentifications (FP) and missed identifications (FN).
Recurring FP and FN cases were then examined to develop a taxonomy of failure modes, as shown in Fig.~\ref{fig:taxonomy}.
Misidentifications originated from non-X-ray diffraction patterns, generic peak-like plots, axis/label confusion, and 2D maps with diffraction-like visual features, while missed identifications occurred due to non-traditional XRD formats, low-resolution XRD, map-style diffraction, and PDF formatting issues.

Across all models, FPs were more frequent than FNs (208 vs. 92), indicating that misidentifying non-XRD figures was a greater challenge than missing true XRD figures.
These differences are important for database construction because FPs can introduce false information into the resulting database and increase API costs, whereas FNs primarily reduce coverage. 
The prevalence and type of these failure modes, however, varied substantially across models (Fig.~\ref{fig:failure_distribution}). 
\texttt{GPT-5.2} produced the fewest FPs, all in one mode (non-X-ray diffraction), whereas other models produced FPs across all four modes, suggesting that \texttt{GPT-5.2} exhibited a narrower and more specific failure behavior.
\texttt{Claude Sonnet~4.5} was better able to distinguish non-XRD diffraction figures, but its failures arose mostly from axis/label confusion, indicating that axes and labels can mislead the model.
\texttt{Gemini~2.5 Flash} and \texttt{Grok~4.1} showed broader failure behavior, indicating greater sensitivity to both visual similarity and contextual ambiguity.

For FNs, \texttt{Gemini~2.5 Flash} missed the fewest true XRD figures, consistent with its highest recall, whereas the other models produced more FNs, though still fewer than their FPs.
For scientific database construction frameworks like \xrdreader, FPs are more critical because they can introduce incorrect information, whereas FNs only reduce coverage by leaving true records absent.
Considering both FPs and FNs, \texttt{GPT-5.2} showed the strongest figure-identification performance, with the fewest total failures. 
These results show that model selection should consider both overall performance and the types of failures that can affect the resulting database. 
One potential way to reduce these failures is through few-shot~\cite{brown2020language} and chain-of-thought prompting~\cite{wei2022chain}.

\phantomsection
\textbf{XRD metadata extraction failure modes.}\label{subsec:xrd_metadata_extraction_failure}
We next examined the metadata failures from manual assessment of the same 47 publications.
Of the 443 scored fields (Fig.~\ref{fig:metadata_perfield}), 41 did not match the curated reference: 35 were missed and 6 were extracted incorrectly.
Lattice parameters accounted for 10 of the 35 misses, followed by radiation type (5), wavelength (4), Wyckoff positions (4), and X-ray density (4).
The six incorrect cases involved space group, space-group number, radiation type, wavelength, Miller indices, and scan range.
Overall, metadata-extraction failures were dominated by missed rather than incorrect values, consistent with high precision observed in the manual assessment.

In summary, XRD figure identification failures were dominated by misidentifications, whereas metadata-extraction failures were dominated by missed values.

\subsection*{Scalability and model selection}\label{subsec:xrdreader_performance}

Finally, we evaluated the scalability of \xrdreader in terms of runtime, API cost, token usage, and the trade-off between computational resources and extraction accuracy across different LLMs to guide model selection for large-scale runs.

\begin{figure*}
\begin{center}
\includegraphics[scale=0.39]{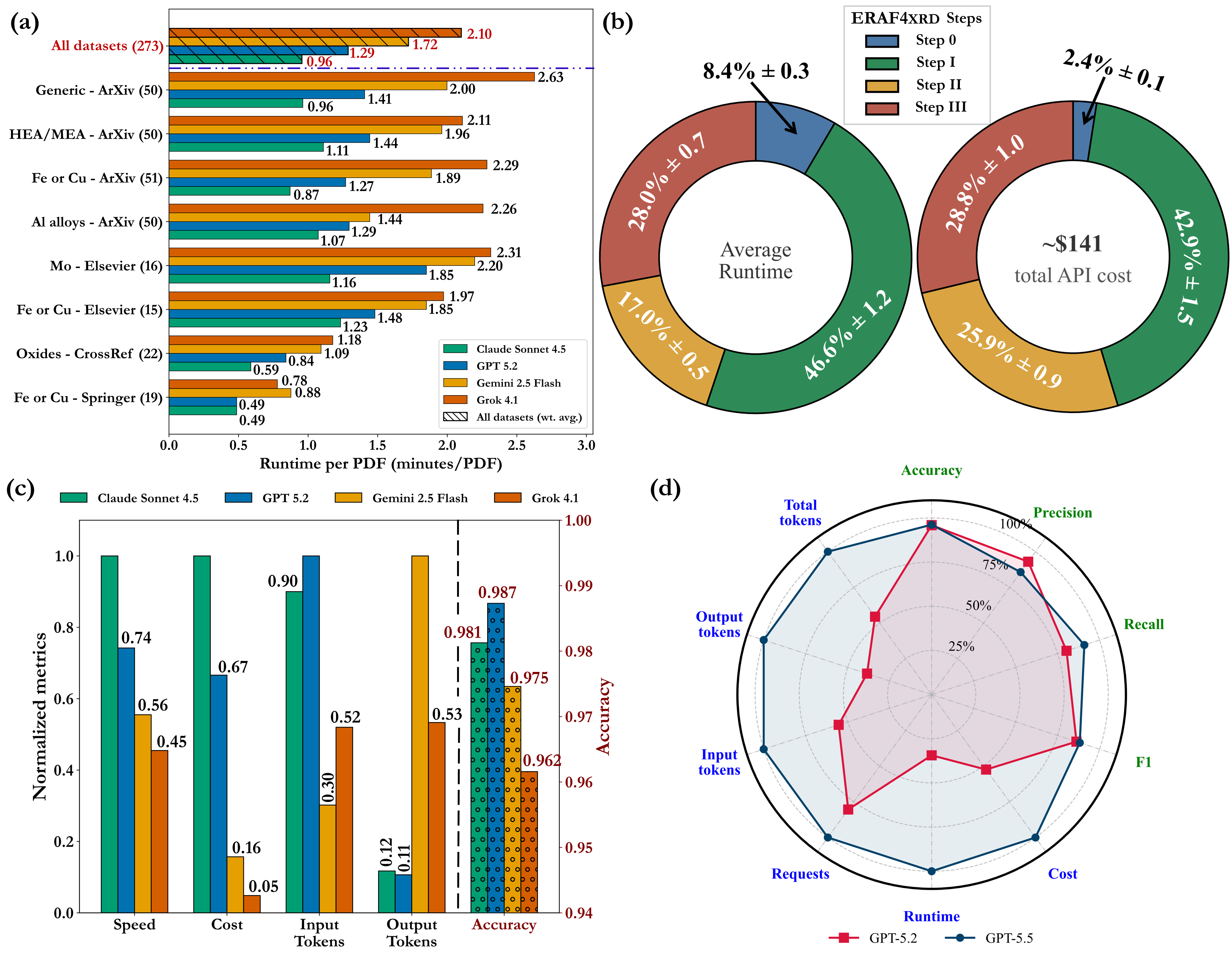}
\caption{\label{fig:scalability}
\xrdreader scalability and model-choice trade-offs. (a) Average runtime per PDF for each LLM across the benchmark datasets; solid bars show dataset-specific runtimes, and hatched bars show the average across all datasets. (b) Step-level contribution to total runtime and total API cost across the 273-publication benchmark. (c) Normalized resource metrics for \texttt{Claude Sonnet~4.5}, \texttt{GPT-5.2}, \texttt{Gemini~2.5 Flash}, and \texttt{Grok~4.1}, with XRD figure identification accuracy (dotted bars) shown on the right axis. (d) \texttt{GPT-5.2} and \texttt{GPT-5.5} comparison on XRD figure identification metrics and resource usage.
}
\end{center}
\end{figure*}

\textbf{Runtime.}
Figure~\ref{fig:scalability}a shows the average runtime per PDF for each LLM across individual datasets (solid bars) and all datasets (hatched bars), with all four agents enabled (Steps~0--III).
\texttt{Claude Sonnet 4.5} was fastest at 0.96~min/PDF, followed by \texttt{GPT-5.2} (1.29~min/PDF), \texttt{Gemini 2.5 Flash} (1.72~min/PDF), and \texttt{Grok 4.1} (2.10~min/PDF).
Thus, the fastest and slowest models differed by more than a factor of two, showing that LLM choice directly affects \xrdreader throughput.
Further details are provided in Supplementary Discussion 6.

\textbf{API-cost and runtime bottlenecks.}\label{subsec:runtime_by_step}
Across all steps, XRD figure identification and figure-level metadata extraction (Step~I) were the dominant computational bottlenecks, accounting for 46.6\% of runtime and 42.9\% of API cost due to candidate figure extraction, pre-filtering, agentic XRD figure identification, figure-level metadata extraction, and initial JSON construction (see Methods).
Validation (Step III) was the second-largest contributor, accounting for 28.0\% of runtime and 28.8\% of cost, followed by Step II (17.0\% and 25.9\%) and Step 0 (8.4\% and 2.4\%).
These results further justify the targeted preprocessing and selective use of visual context adopted in \xrdreader, which limits unnecessary LLM processing during figure identification and data--metadata extraction.
Thus, future runtime improvements should primarily target Step~I, especially through more efficient figure filtering, batching, and parallelization.

\textbf{Resource--accuracy trade-offs.}
Figure~\ref{fig:scalability}c compares runtime, API cost, token usage, and accuracy across the four LLMs.
No model was optimal across all metrics.
\texttt{Claude Sonnet 4.5} was fastest but had the highest API cost, indicating a speed--cost trade-off.
\texttt{GPT-5.2} provided the best overall balance, with the highest XRD figure identification accuracy (98.7\%) and the second-fastest runtime.
\texttt{Gemini 2.5 Flash} required more tokens, whereas \texttt{Grok 4.1} was slower and less efficient despite comparable accuracy.

\textbf{Does a newer LLM improve performance?} On the Generic--ArXiv dataset, \texttt{GPT-5.5} provided only a marginal improvement over \texttt{GPT-5.2}, increasing XRD figure identification accuracy from 95.9\% to 96.2\%, with no substantial F1-score gain (Fig.~\ref{fig:scalability}d). However, this small improvement came with substantially higher computational costs, showing that newer models do not necessarily provide proportional gains.
This emphasizes the importance of benchmarking model performance against computational cost rather than selecting models based solely on capability or recency.

\textbf{LLM selection for \xrdreader.}
Across all tested models, \texttt{GPT-5.2} provided the best overall balance of accuracy, precision, recall, runtime, token usage, and API cost.
\texttt{Claude Sonnet~4.5} is preferable when speed is the highest priority and higher API cost is acceptable.
\texttt{Gemini~2.5 Flash} can be useful when recall is prioritized, but its high FP behavior requires additional caution and scrutiny, especially for database construction, where false XRD records or metadata are more detrimental than missing entries.
\texttt{Grok~4.1} is suitable only for low-cost, low-throughput runs where reduced reliability is acceptable.
Newer frontier models do not necessarily provide sufficient improvement to justify their additional computational cost.

\section*{Discussion}\label{sec:discussion}

A key bottleneck in AI-enabled experimental science is the lack of scalable pathways for converting information embedded in scientific publications into large, realistic, metadata-rich experimental datasets. Decades of publications contain valuable experimental data and expert information, but this information remains fragmented across plots, captions, methods sections, and tables. \xrdreader addresses this challenge for XRD by integrating figure identification, data--metadata linking, and source-evidence-based validation. 
The key challenge is not simply extracting metadata, but correctly linking it to the corresponding experimental data and validating those relationships. 
\xrdreader therefore moves beyond information extraction toward the automated reconstruction of validated experimental records. 
Although demonstrated here for XRD, \xrdreader provides a general framework for transforming literature-embedded experimental information into structured datasets ready for AI and data-driven science.

Independent validation emerges as a critical component of reliable LLM-based database construction. In \xrdreader, source-evidence-based validation modified 44.5\% of the initially extracted metadata, yet the final records achieved 98.5\% precision and 90.7\% recall, with no unsupported metadata observed during manual assessment.
This contrast shows that \xrdreader's strong final performance depends on the automatic, independent validation of extracted data and data--metadata linking.
Combining source-evidence-based LLM validation with non-LLM deterministic scientific checks, therefore, provides a general automated strategy for converting LLM-extracted information into reliable database records, thereby reducing human intervention at scale.

More broadly, our results highlight that reliability and completeness are distinct objectives in automated scientific database construction.
Maximizing extracted information is not worthy if the extracted values or their links to experimental data cannot be trusted. 
\xrdreader therefore prioritizes source-supported information and validated data--metadata links over maximum completeness. 
This design favors a database in which retained records have strong source evidence, while allowing uncertain or unextracted information to remain missing rather than introducing unsupported records. 
This approach is particularly important for experimental databases intended for quantitative analysis and machine learning, where incorrect individual records can propagate into erroneous downstream scientific conclusions.

Benchmarking extraction and reconstruction strategies shows that more LLM input or more complex reasoning does not necessarily improve performance.
Targeted preprocessing of XRD-relevant content was more accurate and efficient than providing the full document, while agentic screening improved performance for smaller models but provided little benefit when a stronger model was used. 
For figure identification and data--metadata linking, however, combining the figure crop with the full-page view improved performance because the two inputs provide complementary information. 
Similarly, combining regex-based metadata extraction with LLM validation provided higher precision than direct LLM extraction. 
Together, these results show that \xrdreader benefits from selectively combining LLM reasoning, multimodal context, and deterministic methods tailored to each task's requirements, rather than relying on more context or more capable models alone.

Scalability similarly depends on how computational resources are allocated across the framework, rather than simply on using more capable LLMs.
Runtime and API cost varied substantially among models, and figure identification and figure-level extraction remained the largest computational bottleneck in the framework. 
This further supports the targeted preprocessing and selective visual inputs used in \xrdreader and identifies opportunities for improvement through better figure filtering, batching, and parallelization. 
\texttt{GPT-5.2} provided the strongest balance of identification performance and computational efficiency among the tested models, whereas different models favored speed, recall, or other resource considerations. 
Moreover, the newer \texttt{GPT-5.5} model produced only a marginal improvement in identification accuracy on the tested dataset despite substantially greater computational cost. 
Thus, literature-scale scientific extraction should not assume that the newest or largest available model is necessarily the most appropriate.

Several areas for improvement remain. 
Unconventional XRD representations and PDF-formatting issues can lead to missed figures, while less common metadata formats remain challenging to extract consistently. 
More flexible extraction strategies could improve completeness while preserving the high precision achieved here. The low-cost CIF-based assessment should be expanded to larger and more diverse material systems.
\xrdreader also currently reconstructs validated figure-specific metadata but does not fully digitize numerical diffraction data from figures.
Combining these validated records with reliable plot digitization is an important next step toward fully machine-readable, literature-derived diffraction datasets.
Preliminary applications to other experimental domains, such as Raman and EXAFS, suggest that the underlying \xrdreader framework can be extended beyond XRD through domain-specific configurations.

\section*{Methods}\label{sec:methods}

\renewcommand{\labelenumi}{(\alph{enumi})}
\subsubsection*{Step 0: Document collection and agentic screening} 
Step 0 performs two main functions: \textit{Document collection}, which collects XRD-related PDF documents through automated download or manual upload, and \textit{Agentic screening}, which determines whether each collected document should be \texttt{kept} or \texttt{rejected} for later steps based on the presence of XRD content.

\paragraph{Document collection:} The automated pipeline downloads open-access and CC-licensed PDF documents from multiple publication sources through application programming interfaces (APIs) using user-defined keyword queries.
Each query consists of two components: `\textit{Element}' and `\textit{Technique}'.
For example, a query for Cu-related XRD articles can be written as \{(`\textit{Cu}' OR `\textit{Copper}') AND (`\textit{X-ray diffraction}' OR `\textit{XRD}')\}.
For material-independent searches, the `\textit{Element}' field can be left empty, as done to retrieve the Generic--ArXiv benchmark dataset used in this study.
\xrdreader currently supports automated collection from ArXiv, Elsevier, Springer, and CrossRef.
During document collection, \xrdreader checks whether files are valid and readable, removes incomplete or preview-only files, and prevents duplicate downloads by checking existing files.
Alternatively, users can manually upload PDFs to bypass this API-based automated download pipeline.

\paragraph{Agentic screening:} 
After collection, \xrdreader uses an LLM-based agent to screen each document for XRD relevance and decide whether to \texttt{keep} or \texttt{reject} it for downstream processing.
This screening serves as an early quality-control filter, preventing documents that mention XRD only in passing—without providing diffraction plots and metadata—from incurring unnecessary API costs in later resource-intensive extraction steps.
The screening agent operates within a decision budget (default $N$=$6$), which defines the maximum number of ReAct-style \cite{yao2022react} iterations (\textit{i.e.,} observe--reason--act--repeat) allowed per document.
During each iteration, the agent gathers evidence through predefined tools that support the screening decision.
These agent tools include retrieving the abstract, searching the PDF, accessing text from a specific page, and inspecting the table of contents.
This observe--reason--act loop continues until the LLM calls \texttt{finalize} with a \texttt{keep} or \texttt{reject} decision or the budget $N$ is exhausted. If the budget is exhausted without a \texttt{finalize} call, the document is rejected; the budget and available tool set can be expanded when more extensive screening is needed.
For each document, the agent stores the final decision, confidence score, reasoning, and complete tool-call trace, making the screening process auditable. To avoid repeated screening, \xrdreader caches results based on the file contents, LLM, and decision budget; a document is re-screened only when the model or budget changes.


\subsubsection*{Step I: XRD figure identification, figure-level metadata extraction, and JSON construction}
After Step~0 screening, each retained document is passed to Step~I, which performs three major tasks: XRD figure identification and storage, figure-level XRD metadata extraction, and JSON construction.

\paragraph{XRD figure identification and storage:}
\label{subsubsec:stepI_figure_extraction}
XRD figure identification consists of three sequential subtasks: candidate figure extraction, candidate figure pre-filtering, and LLM-based XRD figure identification.
Step~I first uses Python-based tools to extract candidate figures from each PDF through a dual raster--vector strategy.
This strategy accounts for the different ways scientific figures are encoded in PDFs: raster figures are extracted directly from embedded image objects, whereas vector figures are reconstructed by grouping the drawing primitives that form the plot.
The extracted candidates can include XRD figures, non-XRD figures, and non-figure entities such as journal logos and watermarks. 
Before LLM inspection, Step~I therefore applies a lightweight non-LLM pre-filter based on candidate type (raster, vector, or mixed), area, extracted file size, and proximity to XRD-relevant text. 
This pre-filter removes non-figure entities and reduces the number of candidates sent to the LLM, thereby lowering API cost. 

The remaining candidates are examined by a vision-capable ReAct-style agent \cite{yao2022react}, which determines whether (a) the candidate is a figure or non-figure entity, (b) it contains an XRD plot, and (c) the XRD data are experimental, simulated, or both.
The agent receives two complementary visual inputs: a high-resolution crop of the candidate figure and a full-page view highlighting its location.
The candidate crop supports detailed inspection of plot features, whereas the full-page view provides contextual information needed to interpret the figure and establish data--metadata links.
When the initial visual inputs are insufficient, the agent can use predefined tools to retrieve nearby or page-level text, inspect adjacent pages and captions, or expand an incomplete candidate crop.
Each confirmed XRD plot is then saved as an image file in a dedicated per-document output folder.
Further implementation details of candidate figure extraction, pre-filtering, and LLM-based XRD figure identification are provided in Supplementary Note 1.

\paragraph{Figure-level XRD metadata extraction:}
For each confirmed XRD figure, the agent extracts metadata directly available from the figure itself.
These figure-level metadata include axis labels, legends, materials, phases, peak positions, and Miller indices, together with the figure and page numbers for tracking.
The Step~I metadata are restricted to information contained in the figure and later complement the text-derived metadata extracted in Step~II.

\paragraph{JSON construction:}
Step~I organizes the extracted information into a per-document JSON object.
For each confirmed XRD figure, the JSON stores the figure-identification decisions, confidence scores, figure and page numbers, extracted figure-level metadata, and the associated agent decision record and tool-call trace.
Paper-level information, including the title and authors, is also stored.
The resulting JSON object serves as the input to Step~II.

\subsubsection*{Step II: Text-derived XRD metadata enrichment and data--metadata linking}
Step~I populates the JSON file with metadata extracted directly from the identified XRD plots; however, complementary XRD metadata may also be reported in the document text.
Step~II extracts this text-derived information and links it to the appropriate XRD data.
Step~II consists of three tasks: text-derived candidate metadata and evidence extraction, data--metadata linking and traceability, and JSON cleaning.

\paragraph{Text-derived candidate metadata and evidence extraction:}
Step~II first assembles candidate text-derived XRD metadata and supporting source evidence using complementary global- and figure-level operations.
The global-level operation is applied once per document and uses predefined regex patterns to identify candidate XRD metadata values from the full document text.
Because regex matching alone cannot determine whether a value applies globally or to a specific XRD figure, these values are treated as provisional candidates.
The same operation also constructs an XRD-focused evidence corpus from diffraction-related text in the document, which serves as the primary evidence for the agent for evaluating these metadata candidates.
The figure-level operation is then applied to each confirmed XRD figure to assemble figure-specific evidence, including its caption, nearby and page-level text, and Step~I-obtained figure-level metadata.
This local context is particularly important when a document reports multiple values for the same metadata field.
When the assembled evidence is insufficient, the Step~II agent can use tools to retrieve additional information from the document.

\paragraph{Data--metadata linking and traceability:}
After the global and figure-level evidence has been assembled, a ReAct-style agent \cite{yao2022react} evaluates each candidate metadata value against the source evidence.
Metadata that apply to the overall XRD methodology are stored in the global metadata block, whereas figure-specific values are stored in the corresponding figure record and linked to the corresponding identified XRD figure record.
Here, in the term `data--metadata links', `data' refers to the identified XRD figure record, not numerical intensity--$2\theta$ data.
For example, if a document reports Cu~K$\alpha$ radiation for powder XRD and Mo~K$\alpha_1$ radiation for single-crystal measurements, the figure-specific context is used to determine which radiation source applies to each XRD figure.
For each retained metadata field, \xrdreader records the supporting source evidence and extraction pathway, making the resulting data--metadata links traceable.

\paragraph{JSON cleaning:}
After enrichment, the JSON contains the extracted XRD metadata together with intermediate agent traces and decision records.
Step~II removes these agent-related fields, including tool-call logs, collected evidence, and decision records, before Step~III validation.
This prevents the validation agent from relying on prior agent reasoning while producing a more compact JSON containing only paper-identification information and the retained global and figure-level XRD metadata.
An example of the JSON before and after cleaning is shown in Supplementary Fig.~1.
Further implementation details of the three Step~II tasks are provided in Supplementary Note 2.

Together, Steps~I and~II form a multimodal extraction approach that combines figure- and text-derived metadata, as discussed in Supplementary Note 3 and illustrated in Supplementary Fig.~2.

\subsubsection*{Step III: Independent validation of XRD figures, metadata, and data--metadata links}
\label{subsec:stepIII_independent_validation}
After Step~II, the cleaned JSON file is passed to Step~III for independent validation of the extracted XRD figures, metadata, and data--metadata links.
Here, ``independent'' means that the Step~III agent has no access to prior agent traces and must re-establish support directly from the source document.
Step~III can also use a different LLM from the preceding steps, providing an additional layer of independence.
Validation proceeds through two sequential processes: LLM-based validation and non-LLM deterministic consistency checks.

\paragraph{LLM-based validation:}\label{PIII_ag}
Step~III first uses a ReAct-style agent \cite{yao2022react} to independently validate the global and figure-level XRD metadata blocks.
The agent operates within a default tool-call budget of $N=12$ for each metadata block and retrieves textual and visual evidence directly from the source document.
Textual evidence can be obtained from methods-section text, targeted document searches, page- and figure-level context, tables, captions, and references, whereas visual evidence is obtained from the Step~I figure crop and full-page view.
This multimodal evidence allows the agent to evaluate metadata values and their contextual relationships to the extracted XRD figures.

For each metadata field, the agent assigns a status of \texttt{supported}, \texttt{unsupported}, or \texttt{unclear} based on the gathered source evidence. 
A metadata field is marked \texttt{supported} when the gathered evidence confirms the value, \texttt{unsupported} when the evidence contradicts or provides no support for it, and \texttt{unclear} when related evidence is available but is insufficient to confirm or reject the value.
The agent also checks whether each metadata value is correctly assigned to the global or corresponding figure-level metadata block and, for figure-level values, whether its data--metadata link to the XRD figure is correct.
Step~III can retain, correct, remove, or, when directly supported by the source, add new metadata values.
Therefore, Step~III acts both as a validation and enhancement step.

To validate the XRD figures without repeating the costly figure-identification procedure, Step~III uses a simple metadata-based strategy.
It first examines the validated figure-level metadata block for each figure. 
A figure is retained if this block contains at least one validated metadata value.
If the figure-level block is empty, Step~III then checks whether the validated global metadata apply to that figure. 
If they do, the figure is retained; otherwise, it is flagged as a potential misidentification for further review.
This approach validates the identified XRD figures without requiring a second full LLM-based XRD figure identification.
Further details are provided in Supplementary Note 4.

\xrdreader also provides an optional lower-cost, non-agentic validation mode in which the pre-assembled textual evidence is sent to the LLM in one API call.
This mode reduces API calls and runtime but does not use visual evidence and is therefore less suitable when data--metadata linking depends strongly on visual context, as is often the case for XRD.
In contrast, it may be more suitable for predominantly text-based links, such as linking a material to a reported property.
Therefore, agentic multimodal validation is used as the default mode for XRD.

\paragraph{Non-LLM deterministic checks:}
After LLM-based validation, Step~III applies a final set of non-LLM deterministic checks to identify crystallographic or XRD methodological inconsistencies.
These checks are needed because \xrdreader treats the source document as the ground truth for as-reported metadata; therefore, a value can be correctly extracted from the source yet still be inconsistent with basic crystallographic or radiation-source rules.
Step~III checks three consistency relationships: space group--crystal system, lattice-parameter--crystal-system, and radiation source--wavelength.
For example, Fm$\overline{3}$m paired with a tetragonal crystal system, or Cu~K$\alpha$ radiation paired with a wavelength of 0.7107~\AA, is flagged as inconsistent.
When an inconsistency is detected, the field is downgraded from \texttt{supported} to \texttt{unsupported}, and the inconsistency is recorded for expert review.
Because these checks are rule-based and always run after agentic validation, they provide an additional validation layer without increasing API cost.
Further implementation details are provided in Supplementary Note 4.

\backmatter

\bmhead{Supplementary information}
A supplementary information file is available with this article.

\bmhead{Competing interests}
All authors declare no financial or non-financial competing interests. 

\bmhead{Author contributions}
A.M. developed the framework, implemented the software, constructed and curated the benchmark dataset, performed the computational experiments and benchmarking, analyzed the results, and contributed to writing and editing of the manuscript. S.J.L.B. extended A.M.'s initial framework development and contributed to the software architecture and packaging of the framework as a scikit package. W.R. and S.J.L.B. contributed to the overall framework design, interpretation of the results, and revision of the manuscript. N.A. conceived the study, supervised and administered the project, contributed to the framework design, benchmark dataset curation, and interpretation of the results, and revised and finalized the manuscript. All authors reviewed and approved the manuscript.

\bmhead{Data availability}
The benchmark dataset, ground truth data, \xrdreader outputs, and analysis spreadsheets supporting the findings of this study are available in the following Zenodo link: \href{https://zenodo.org/records/22683615?preview=1&token=eyJhbGciOiJIUzUxMiJ9.eyJpZCI6ImRkZDQ1NDk3LWQxYTAtNDkwYi05ZTIxLTE1YzQxMDYxYjc5ZSIsImRhdGEiOnt9LCJyYW5kb20iOiI1Zjc0ZjMyMTczOGQyMDA4Mzg5NDUyZGMyNDcyMTQ1OSJ9.slcub0nANZZuytN-JtrdxjPdW33_D--hE6Mo7QG1IyShQe1mRk4Q5qimWdy8e767g9UApdg4gVlUm7vCOhuyDw}{https://doi.org/10.5281/zenodo.22683615}

\bmhead{Code availability}
The source code for \xrdreader is openly available at \href{https://github.com/niaz60/ERAF4XRD}{github.com/niaz60/ERAF4XRD}.



\bmhead{Acknowledgements}
This work was supported by the U.S. National Science Foundation (NSF), Office of Advanced Cyberinfrastructure (OAC), under Award Nos. 2513797 and 2625202.
Unless otherwise noted, all NIST work was funded solely by the U.S. government.
A.M. acknowledges the use of ChatGPT and Gemini for language editing, rephrasing, and grammar checking.
A.M. also acknowledges the use of Codex and Claude Code for assistance with framework development and code debugging.
However, all scientific ideas, analyses, interpretations, tables, figures, and final manuscript content are the responsibility of the authors. 

\bmhead{Declarations}
We identify certain commercial equipment, instruments, and materials in this article to specify adequately the experimental procedures. 
In no case does such identification imply recommendation or endorsement by the National Institute of Standards and Technology nor does it imply that the materials or equipment identified are necessarily the best available for the purpose.


\end{document}